\documentclass[conference]{IEEEtran}

\usepackage{amsfonts,amsmath,amssymb}
\usepackage{dsfont}
\usepackage{cite}
\usepackage{citesort}
\usepackage{graphicx}
\usepackage{url}
\usepackage{caption}
\usepackage{bm}
\usepackage{bbding}
\usepackage{epstopdf}
\usepackage{algorithm}
\usepackage{algorithmic}
\usepackage{CJK}

\DeclareMathOperator*{\argmin}{argmin}
\newtheorem{theorem}{\bf{Theorem}}

\newtheorem{proposition}{\bf{Proposition}}

\makeatletter
\def\ScaleIfNeeded{%
\ifdim\Gin@nat@width>\linewidth \linewidth \else \Gin@nat@width
\fi } \makeatother

\title{Location-Based Beamforming for Rician Wiretap Channels}

\begin{document}

\author{\IEEEauthorblockN{Chenxi Liu and
Robert Malaney}
\IEEEauthorblockA{School of Electrical
Engineering and Telecommunications, The University of New South
Wales, Sydney, Australia}
Email: chenxi.liu@student.unsw.edu.au, r.malaney@unsw.edu.au}

\maketitle

\begin{abstract}
We propose a location-based beamforming scheme for wiretap channels, where a source communicates with a legitimate receiver in the presence of an eavesdropper. We assume that the source and the eavesdropper are equipped with multiple antennas, while the legitimate receiver is equipped with a single antenna. We also assume that all channels are in a Rician fading environment, the channel state information from the legitimate receiver is perfectly known at the source, and that the only information on the eavesdropper available at the source is her location.
We first describe how the beamforming vector that minimizes the secrecy outage probability of the system is obtained, illustrating its dependence on the eavesdropper's location. We then derive an easy-to-compute expression for the secrecy outage probability when our proposed location-based beamforming is adopted. Finally, we investigate the impact location uncertainty has on the secrecy outage probability, showing how our proposed solution can still allow for secrecy even when the source has limited information on the eavesdropper's location.
\end{abstract}
\section{Introduction}
Physical layer security has attracted significant research attention recently. Compared to the traditional upper-layer cryptographic techniques using secret keys, physical layer security safeguards wireless communications by directly exploiting the randomness offered by wireless channels without using secret keys, and thus has been recognized as an alternative for cryptographic techniques \cite{Yang_Mag}. The principle of physical layer security was first studied in \cite{wyner} assuming single-input single-output systems. It was shown that secrecy can only exist when the wiretap channel between the source and the eavesdropper is a degraded version of the main channel between the source and the legitimate receiver. Subsequently, this result was generalized to the case where the main channel and the wiretap channel are independent \cite{csiszar}.

Most recently, implementing multi-input multi-output (MIMO) techniques at the source/legimitmate receiver has been shown to significantly improve the physical layer security of wiretap channels \cite{khisti,wornell,chenxi,chenxi2,nan4,Zhang13,nan5,nan6,nan,nan3,jiangyuan}. In terms of MIMO techniques, beamforming \cite{khisti,wornell,chenxi,chenxi2,nan4,jiangyuan}, artificial noise (AN) \cite{Zhang13,nan5,nan6}, and transmit antenna selection \cite{nan,nan3} are just a few techniques that can be utilized to boost the physical layer security of wiretap channels. In \cite{khisti,wornell,chenxi,chenxi2,nan4,Zhang13,nan5,nan6,nan,nan3,jiangyuan}, it is assumed that the channel state information (CSI) from the eavesdropper is perfectly or statistically known at the source. This assumption, however, is unlikely to be valid in practice - especially when the eavesdropper is not an authorized component of the communication system.

In this paper we propose a location-based beamforming scheme  that does not require any form of CSI be passed by the eavesdropper back to the source. Rather, we will assume that some \emph{a priori} known location information of the eavesdropper is available to the source. Such a scenario can occur in many circumstances, such as those detailed in \cite{shihao_axiv}. In our scheme, we assume  that \emph{all} of the communication channels are in a Rician fading environment. That is, the channel between the source and the legitimate receiver and the channel between the source and the eavesdropper are a combination of a line-of-sight (LOS) component \emph{and} a random scattered component.  We also assume that the CSI from the legitimate receiver is \emph{perfectly} known at the source, while the \emph{only} information on the eavesdropper available at the source is her location. Our key goal is to determine the beamforming vector at the source that minimizes the secrecy outage probability of the system, given the CSI of the main channel and the eavesdropper's location.


Perhaps the most relevant work to ours is that of  \cite{shihao_axiv} in which the secrecy outage probability in Rician wiretap channels was investigated, largely for the case where the location of the eavesdropper was available at the source but where the CSI of the main channel was unavailable.
Compared to \cite{shihao_axiv}, our work is different in the following main aspects:
(i) We derive a simpler expression of the secrecy outage probability  when the eavesdropper's location and the  CSI of the main channel are known. We highlight that our  expression is valid for arbitrary values of average signal-to-noise ratios (SNR) and Rician $K$ factors in the main channel and the wiretap channel.
(ii) Based on this new expression we develop a much more efficient search algorithm for the determination of the optimal beamforming scheme that minimizes the secrecy outage probability when the  CSI of the main channel and the eavesdropper's location are available at the source. We highlight that our new search algorithm invokes a one-dimensional search, as opposed to the multi-dimensional searches required previously, thereby greatly reducing the computational complexity (important for in-field deployments).
(iii) We examine the impact of location uncertainty on the secrecy outage probability, showing how  secrecy can still exist when only a noisy estimate of the eavesdropper's location is available at the source.


\section{System Model}
\begin{figure}[t]
\begin{center}{\includegraphics[width=0.8\columnwidth]{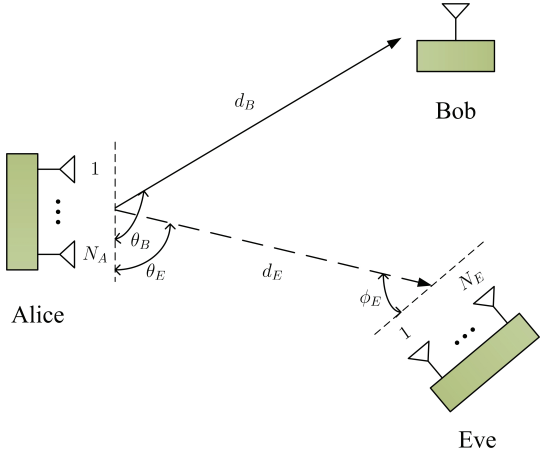}}
\caption{Illustration of a wiretap channel with Rician fading.}\label{system_model}
\end{center}
\end{figure}

We consider a wiretap channel with Rician fading, as depicted in Fig. \ref{system_model}, where Alice communicates with Bob in the presence of Eve (the eavesdropper). In this channel, Alice and Eve are equipped with uniform linear arrays (ULA) with $N_A$ and $N_E$ antennas, respectively, while Bob is equipped with a single antenna.
We adopt the polar coordinate system. As such, the locations of Alice, Bob, and Eve are denoted by $\left(0,0\right)$, $\left(d_B,\theta_B\right)$, and $\left(d_E,\theta_E\right)$, respectively. We consider that the main channel between Alice and Bob and the eavesdropper's channel between Alice and Eve are subject to quasi-static independent and identically distributed (i.i.d) Rician fading with different Rician $K$-factors. We also consider that a $K$-factor map ($K$ as a function of location) is known in the vicinity of Alice via some {\em a priori} measurement campaign.
We assume that the CSI of the main channel is known to Alice, while the only available information on Eve is her location. This assumption is reasonable in some military application scenarios where Alice can obtain Eve's
location through some {\em a priori} surveillance.

We denote $\mathbf{h}$ as the $1\times N_A$ channel vector from Alice to Bob, which is given by
\begin{align}
\label{bob_channel} \mathbf{h} =\sqrt{\frac{K_B}{1+K_B}}\mathbf{h}_{o} + \sqrt{\frac{1}{1+K_B}}\mathbf{h}_{r},
\end{align}
where $K_B$ denotes the Rician $K$-factor of the main channel, $\mathbf{h}_{o}$ denotes the LOS component, and $\mathbf{h}_r$ denotes the scattered component - the elements of which are assumed to be i.i.d complex Gaussian random variables with zero mean and unit variance, i.e., $\mathbf{h}_r\sim\mathcal{CN}\left(\mathbf{0}_{N_A},\mathbf{I}_{N_A}\right)$. In \eqref{bob_channel}, $\mathbf{h}_o$ is expressed as \cite{Tsai}
\begin{align}
\label{h_o} \mathbf{h}_o = \begin{bmatrix}
  1,\cdots,\exp\left(j2\pi\left(N_A-1\right)\delta_A\cos\theta_B\right)
\end{bmatrix},
\end{align}
where $\delta_A$ denotes the constant spacing, in wavelengths, between adjacent antennas of the ULA at Alice. We also denote $\mathbf{G}$ as the $N_E\times N_A$ channel matrix from Alice to Eve, which is given by
\begin{align}
\label{eve_channel} \mathbf{G} = \sqrt{\frac{K_E}{1+K_E}}\mathbf{G}_{o} + \sqrt{\frac{1}{1+K_E}}\mathbf{G}_{r},
\end{align}
where $K_E$ denotes the Rician $K$-factor of the eavesdropper's channel, $\mathbf{G}_{o}$ denotes the LOS component, and $\mathbf{G}_{r}$ denotes the scattered component - the elements of which are assumed to be i.i.d complex Gaussian random variables with zero mean and unit variance, i.e., $\mathbf{G}_r\sim\mathcal{CN}\left(\mathbf{0}_{N_E},\mathbf{I}_{N_E}\right)$. In \eqref{eve_channel}, $\mathbf{G}_o$ is expressed as \cite{Taricco}
\begin{align}
\label{G_o} \mathbf{G}_o = \mathbf{r}_o^T\mathbf{g}_o,
\end{align}
where $\mathbf{r}_o$ denotes the array responses at Eve, which is given by
\begin{align}
\label{r_o} \mathbf{r}_o = \begin{bmatrix}
1,\cdots,\exp\left(-j2\pi\left(N_E-1\right)\delta_E\cos\phi_E\right)
\end{bmatrix},
\end{align}
where $\delta_E$ denotes the constant spacing, in wavelengths, between adjacent antennas of the ULA at Eve, and $\phi_E$ denotes the angle of arrival from Eve to Alice (see Fig.~\ref{system_model}),
and $\mathbf{g}_o$ denotes the array response at Alice, which is given by
\begin{align}
\label{g_o} \mathbf{g}_o = \begin{bmatrix}
1,\cdots,\exp\left(j2\pi\left(N_A-1\right)\delta_A\cos\theta_E\right)
\end{bmatrix}.
\end{align}

According to \eqref{bob_channel}--\eqref{g_o}, we express the received signal at Bob as
\begin{align}
\label{y_b} y_B = \sqrt{P_Ad_B^{-\eta}}\mathbf{h}\mathbf{x}_A + n_B,
\end{align}
where $P_A$ denotes the transmit power at Alice, $\eta$ denotes the path loss component, $\mathbf{x}_A$ denotes the transmitted signal by Alice, and $n_B$ denotes the thermal noise at Bob - which is assumed to be a complex Gaussian random variable with zero mean and variance $\sigma_B^2$, i.e., $n_B\sim\mathcal{CN}\left(0,\sigma_B^2\right)$. In \eqref{y_b}, $\mathbf{x}_A$ is expressed as
\begin{align}
\label{x_a} \mathbf{x}_A = \mathbf{w}t_A,
\end{align}
where $\mathbf{w}$ denotes the $1\times N_A$ beamforming matrix, and $t_A$ is a scalar, which denotes the information signal transmitted by Alice. We assume that $\|\mathbf{w}\|^2=1$ and $\mathbb{E}\left[t_A^2\right]=1$. We then express the received signal at Eve as
\begin{align}
\label{y_e} \mathbf{y}_E = \sqrt{P_Ad_E^{-\eta}}\mathbf{G}\mathbf{x}_A + \mathbf{n}_E,
\end{align}
where $\mathbf{n}_E$ denotes the thermal noise vector at Eve - the elements of which are assumed to be i.i.d complex Gaussian random variables with zero mean and variance $\sigma_E^2$, i.e., $\mathbf{n}_E\sim\left(\mathbf{0}_{N_E},\mathbf{I}_{N_E}\right)$.
As such, we express the received SNR at Bob as
\begin{align}
\label{snr_b} \gamma_B = \overline{\gamma}_B|\mathbf{h}\mathbf{w}|^2,
\end{align}
where $\overline{\gamma}_B={P_Ad_B^{-\eta}}/{\sigma_B^2}$.
 Note, we assume that Eve adopts maximal ratio combining (MRC) \cite{Dighe} to process her received signal (maximizing her SNR). As per the rules of MRC, the received SNR at Eve is expressed as
\begin{align}
\label{snr_e} \gamma_E = \overline{\gamma}_E\|\mathbf{G}\mathbf{w}\|^2,
\end{align}
where $\overline{\gamma}_E={P_Ad_E^{-\eta}}/{\sigma_E^2}$.
\section{Location-Based Beamforming Scheme}
We first describe in detail how the optimal beamforming scheme that minimizes the secrecy outage probability is obtained by utilizing Bob's CSI and Eve's location. We then derive an easy-to-compute expression for the secrecy outage probability when the proposed location-based beamforming scheme is applied.

Based on \eqref{snr_b} and \eqref{snr_e}, the achievable secrecy rate in the wiretap channel is expressed as \cite{Oggier}
\begin{align}
\label{secrecy_rate} C_S = \left\{\begin{array}{ll}
C_B-C_E, &\gamma_B>\gamma_E\\0,&\gamma_B\leq\gamma_E,
\end{array}\right.
\end{align}
where $C_B=\log_2\left(1+\gamma_B\right)$ is the capacity of the main channel, and $C_E=\log_2\left(1+\gamma_E\right)$ is the capacity of the eavesdropper's channel.
In this wiretap channel, if $C_S\geq R_S$, where $R_S$ denotes a given secrecy transmission rate, the perfect secrecy is guaranteed. If $C_S< R_S$, information on the transmitted signal is leaked to Eve, and the secrecy is compromised.  In order to evaluate the secrecy performance of the wiretap channel in detail, we adopt the secrecy outage probability as the performance metric - defined as the probability that the achievable secrecy rate is less than a given secrecy transmission rate conditioned on $\gamma_B$. Mathematically, this is formulated as
\begin{align}
\label{secrecy_outage_probability} P_{\text{out}}\left(R_S\right) = \mbox{Pr}\left(C_S<R_S|\gamma_B\right).
\end{align}
Our goal is to find the optimal beamforming vector that minimizes the secrecy outage probability. That is, we wish to find
\begin{align}
\label{prob_form} \mathbf{w}^{\ast}=\argmin_{\mathbf{w},\|\mathbf{w}\|^2=1} P_{\text{out}}\left(R_S\right).
\end{align}
In order to solve \eqref{prob_form}, we
 present the following proposition.
\begin{proposition}
\label{p1} Given $\tau\in\left[0,1\right]$, the optimal beamforming vector $\mathbf{w}^{\ast}$ that minimizes the secrecy outage probability is a member of the following family of beamformer solutions,
\begin{align}
\label{p1_result} \mathbf{w}\left(\tau\right) = \sqrt{\tau}\mathbf{w}_{\text{ZF}} + \sqrt{1-\tau}\mathbf{w}_{\text{ZF}}^{\bot}.
\end{align}
 Here, $\mathbf{w}_{\text{ZF}} = \frac{\mathbf{\Psi}_{\mathbf{G}_o}^{\bot}\mathbf{h}^H}{\|\mathbf{\Psi}_{\mathbf{G}_o}^{\bot}\mathbf{h}^H\|}$, where $\mathbf{\Psi}_{\mathbf{G}_o}^{\bot}=\mathbf{I}_{N_A}-\mathbf{G}_{o}^H\left(\mathbf{G}_{o}\mathbf{G}_{o}^H\right)^{-1}\mathbf{G}_{o}$; and $\mathbf{w}_{\text{ZF}}^{\bot}= \frac{\mathbf{\Psi}_{\mathbf{G}_o}\mathbf{h}^H}{\|\mathbf{\Psi}_{\mathbf{G}_o}\mathbf{h}^H\|} $ where $\mathbf{\Psi}_{\mathbf{G}_o}=\mathbf{G}_o^H\left(\mathbf{G}_o\mathbf{G}_o^H\right)^{-1}\mathbf{G}_o$.
\begin{proof}
Suppose that $\{\mathbf{w}_{\text{ZF}},\mathbf{w}_{\text{ZF}}^{\bot},\mathbf{w}_3,\cdots,\mathbf{w}_{N_A}\}$ denotes an orthonormal basis in the complex space $\mathbb{C}^{N_A}$. As such, any beamforming vector at Alice can be expressed as \cite{gerbracht}
\begin{align}
\label{p1_proof}\mathbf{w} = \lambda_1\mathbf{w}_{\text{ZF}}+\lambda_2\mathbf{w}_{\text{ZF}}^{\bot}+\sum_{l=3}^{N_A}\lambda_l\mathbf{w}_l,
\end{align}
where $\mathbf{\lambda} = [\lambda_1,\lambda_2,\cdots,\lambda_{N_A}]$ are complex and $\|\mathbf{\lambda}\|^2 = 1$. We first note that the achievable secrecy rate $C_S$ is a function of $\mathbf{w}$. We then note that beamforming into $\mathbf{w}_l$ has no impact on the capacity of the main channel $C_B$. This is due to the fact that $\mathbf{w}_l$ are orthogonal to the plane spanned by $\left\{\mathbf{w}_{\text{ZF}},\mathbf{w}_{\text{ZF}}^{\bot}\right\}$ and the main channel $\mathbf{h}$ lies in this plane. We also find that beamforming into $\mathbf{w}_l$, on the other hand, may increase the capacity of the eavesdropper's channel $C_E$ unless the eavesdropper's channel $\mathbf{G}$ also lies in the plane spanned by $\left\{\mathbf{w}_{\text{ZF}},\mathbf{w}_{\text{ZF}}^{\bot}\right\}$.



Based on the above analysis, we see that beamforming into $\mathbf{w}_l$ decreases $C_S$ or has no impact on $C_S$. As such, we confirm that the optimal beamforming vector has the following structure, given by
\begin{align}
\label{p1_proof_2} \mathbf{w}\left(\tau\right) = \underbrace{{\sqrt{\tau}\exp\left(j\theta_a\right)}}_{\lambda_1}\mathbf{w}_{\text{ZF}} + \underbrace{{\sqrt{1-\tau}\exp\left(j\theta_b\right)}}_{\lambda_2}\mathbf{w}_{\text{ZF}}^{\bot}.
\end{align}
We note that $\left(\theta_a\right)$ and $\left(\theta_b\right)$  in \eqref{p1_proof_2} are general phases have no impact on $C_S$, thus without loss of generality we can set $\theta_a = \theta_b = 0$. Substituting $\theta_a = \theta_b =0$ into \eqref{p1_proof_2} we obtain the desired result in \eqref{p1_result}, which completes the proof.
\end{proof}
\end{proposition}

With the aid of Proposition \ref{p1}, we note that the optimal beamforming vector $\mathbf{w}^{\ast}$ that solves \eqref{prob_form} can be obtained by finding the optimal $\tau^{\ast}$ that minimizes the secrecy outage probability. As such, we re-express \eqref{prob_form} as
\begin{align}
\label{prob_form_2} \tau^{\ast} = \argmin_{0\leq\tau\leq1}P_{\text{out}}\left(R_S\right).
\end{align}
We highlight that Proposition \ref{p1} provides a far more efficient way of obtaining the optimal beamforming vector $\mathbf{w^{\ast}}$ that solves \eqref{prob_form} compared to an exhaustive search. This is due to the fact that an exhaustive search is performed in the complex space $\mathbb{C}^{N_A}$. Consequently, the computational complexity of the exhaustive search grows exponentially as $N_A$ increases. This is to be compared with our method in Proposition \ref{p1} which involves a one-dimensional search of $\tau^{\ast}$ only, regardless of the value of $N_A$.

We now present the expression of the secrecy outage probability when $\mathbf{w}\left(\tau\right)$ is adopted as the beamforming vector in the following theorem.

\begin{theorem}
\label{t1} The secrecy outage probability when $\mathbf{w}\left(\tau\right) = \sqrt{\tau}\mathbf{w}_{\text{ZF}} + \sqrt{1-\tau}\mathbf{w}_{\text{ZF}}^{\bot}$ is adopted as the beamforming vector is given by
\begin{align}
\label{t1_result} P_{\text{out}}\left(R_S\right) = 1- \frac{\gamma\left(N_E\hat{m}_E,\frac{2^{-R_S}\left(1+\gamma_B\right)-1}{\hat{m}_E^{-1}\hat{\overline{\gamma}}_E}\right)}{\Gamma\left(N_E\hat{m}_E\right)},
\end{align}
where $\gamma\left(\cdot,\cdot\right)$ is the lower incomplete gamma function, defined as \cite[Eq. (8.350)]{table},
\begin{align}
\label{lower_incomplete_function} \gamma\left(\mu,\nu\right) = \int_{0}^{\nu}\exp\left(-t\right)t^{\mu-1}dt,
\end{align}
\begin{align}
\label{m_e} \hat{m}_E=\frac{\left(\hat{K}_E+1\right)^2}{2\hat{K}_E+1},
\end{align}
where $\hat{K}_E=|\mathbf{g}_o\mathbf{w}\left(\tau\right)|^2K_E$,
\begin{align}
\label{tilde_gamma_e} \hat{\overline{\gamma}}_E = \mathbb{E}\left[\gamma_E\right] = \frac{\left(K_E|\mathbf{g}_o\mathbf{w}\left(\tau\right)|^2+1\right)\overline{\gamma}_E}{1+K_E},
\end{align}
and $\Gamma\left(\cdot\right)$ is the Gamma function, defined as \cite[Eq. (8.310)]{table},
\begin{align}
\label{gamma_func} \Gamma\left(z\right) = \int_0^{\infty}\exp\left(-t\right)t^{z-1}dt.
\end{align}
\begin{proof}
We  focus on the probability density function (PDF) of $\gamma_E$ when $\mathbf{w}\left(\tau\right)$ is adopted as the beamforming vector,
 which is expressed as \cite{shihao_axiv}
\begin{align}
\label{pdf_snr_e} f_{\gamma_E}\left(\gamma\right) = \left(\frac{\hat{m}_E}{\hat{\overline{\gamma}}_E}\right)^{N_E\hat{m}_E}\frac{\gamma^{N_E\hat{m}_E-1}}{\Gamma\left(N_E\hat{m}_E\right)}\exp\left(-\frac{\hat{m}_E\gamma}{\hat{\overline{\gamma}}_E}\right).
\end{align}
The cumulative distribution function (CDF) of $\gamma_E$ is then obtained as
\begin{align}
\label{CDF_snr_e} F_{\gamma_E}\left(\gamma\right) = \frac{\gamma\left(N_E\hat{m}_E,\frac{\hat{m}_E\gamma}{\hat{\overline{\gamma}}_E}\right)}{\Gamma\left(N_E\hat{m}_E\right)}.
\end{align}
As such, we re-express $P_{\text{out}}\left(R_S\right)$ in \eqref{secrecy_outage_probability} as
\begin{align}
\label{secrecy_outage_probability_2} P_{\text{out}}\left(R_S\right) &= \mbox{Pr}\left(C_B-C_E<R_S|\gamma_B\right)\notag\\
&=\mbox{Pr}\left(C_E>C_B-R_S|\gamma_B\right)\notag\\
&=\mbox{Pr}\left(\gamma_E>2^{-R_S}\left(1+\gamma_B\right)-1\right)\notag\\
&=1-F_{\gamma_E}\left(2^{-R_S}\left(1+\gamma_B\right)-1\right).
\end{align}
Substituting \eqref{CDF_snr_e} into \eqref{secrecy_outage_probability_2}, we obtain the desired result in Theorem \ref{t1}. The proof is completed.
\end{proof}
\end{theorem}

 Note, in Theorem \ref{t1} Eve's location is explicitly expressed in the expressions for $\hat{m}_E$, $\hat{K}_E$, and $\hat{\overline{\gamma}}_E$. Note also, that our derived expression is valid for arbitrary values of average SNRs and Rician $K$ factors in the main channel and the wiretap channel. Based on Proposition \ref{p1} and Theorem \ref{t1}, we see that the optimal $\tau^{\ast}$ that minimizes $P_{\text{out}}\left(R_S\right)$ can be easily obtained through a one-dimensional numerical search.

We point out that $\phi_E$ disappears in the expression for the secrecy outage probability in Theorem \ref{t1}. As an aside, it is perhaps interesting to show why this is so. To this end, we re-express $\gamma_E$ in \eqref{snr_e} as
\begin{align}
\label{snr_e_re} \gamma_E = \overline{\gamma}_E\sum_{i=1}^{N_E}|\mathbf{g}_{i}\mathbf{w}\left(\tau\right)|^2,
\end{align}
where $\mathbf{g}_i$ is the $1\times N_A$ channel vector between Alice and $i$-th Eve's antenna, given by
\begin{align}
\label{g_i} \mathbf{g}_i = \sqrt{\frac{K_E}{1+K_E}}r_{o,i}\mathbf{g}_o + \sqrt{\frac{1}{1+K_E}}\mathbf{g}_{r,i},
\end{align}
where $r_{o,i}$ is the i-th element of $\mathbf{r}_o$, given by $r_{o,i}= \exp\left(-j2\pi\left(i-1\right)\theta_E\cos\phi_E\right)$ and $\mathbf{g}_{r,i}$ is the $i$-th row of $\mathbf{G}_r$. Based on \eqref{g_i}, we express $\mathbf{g}_i\mathbf{w}\left(\tau\right)$ as
\begin{align}
\label{g_i_2} \mathbf{g}_i\mathbf{w}\left(\tau\right) = \sqrt{\frac{K_E}{1+K_E}}r_{o,i}\mathbf{g}_o\mathbf{w}\left(\tau\right) + \sqrt{\frac{1}{1+K_E}}\mathbf{g}_{r,i}\mathbf{w}\left(\tau\right).
\end{align}
We note that $|r_{o,i}\mathbf{g}_o\mathbf{w}\left(\tau\right)|^2 = |\mathbf{g}_o\mathbf{w}\left(\tau\right)|^2$ for any $r_{o,i}$. As such, we confirm that the value of $\phi_E$ has no impact on the secrecy outage probability. This reveals that our analysis reported here is also applicable for antenna arrays other than ULA at Eve.
\section{Impact of Eavesdropper's Location Uncertainty}
 Thus far, we have assumed that Eve's  location is perfectly available at Alice. In this section, we examine the impact of Eve's location uncertainty on the secrecy performance of our proposed location-based beamforming scheme. To this end, we first characterize the uncertainty in Eve's location.

We assume that Eve's location, available at Alice, is obtained through some estimation. This estimation of Eve's location can be made by using received signal strength (RSS), angle of arrival (AOA), time of arrival (TOA), and/or time difference of arrival (TDOA). In addition, we note that there will be errors in the estimated Eve's location due to the noise in the RSS and timing information measurements.
To provide focus, we assume the use of the TDOA scheme, e.g. \cite{4,5}, as the positioning algorithm. Providing such algorithms are close to optimal, we can directly utilize in our analysis a probability distribution of estimated positions derived from the Fisher matrix of the TDOA scheme.

We now detail the Fisher matrix of the TDOA scheme \cite{chenxi4}. For the sake of generality, we assume there exist $N$ anchor points in our system that cooperate to localize Eve. We denote  Eve's true location and the location of the $n^{\text{th}}$ anchor point in a 2-D plane by $\xi_0=[x_0,y_0]$ and $\xi_n=[x_n,y_n]$, respectively.  We  denote the time difference relative to that measured by anchor point $1$ and the $n^{\text{th}}$ anchor point as $\phi_n$, then we obtain the logarithm of the distribution of $\phi_n$ as,
\begin{align}
\label{TDOA} -\ln f\left(\phi_n\right) = \frac{\left(\phi_n-\frac{d_n-d_1}{c}\right)^2}{4c^2\sigma_t^2},
\end{align}
where $c$ denotes the speed of light, $\sigma_t^2$ in the variance of the timings, and $d_n$ denotes the distance between the $n^{\text{th}}$ anchor point and Eve, expressed as
\begin{align}
\label{dn} d_n = \sqrt{\left(x_n-x_0\right)^2 + \left(y_n-y_0\right)^2}.
\end{align}
According to \eqref{TDOA}, we can introduce a variable as $\theta_n = \arctan \frac{y_n-y_0}{x_n-x_0}$, then we express the Fisher matrix of the TDOA scheme as
\begin{align}
\label{fisher_2} \mathbf{J}\left({\phi_n}\right) = \left[ {\begin{array}{*{20}{c}}
   {{J\left({\phi_n}\right)_{11}}} & {{J\left({\phi_n}\right)_{12}}}  \\
   {J\left({\phi_n}\right)_{21}} & {{J\left({\phi_n}\right)_{22}}}  \\
\end{array}} \right],
\end{align}
where
\begin{align}
\label{J_xx_2}J\left({\phi_n}\right)_{11}=\dfrac{1}{2c^2\sigma_t^2}\sum_{n=2}^N\left(\cos\theta_n-\cos\theta_1\right)^2,
\end{align}
\begin{align}
\label{J_yy_2} J\left({\phi_n}\right)_{22}
=\dfrac{1}{2c^2\sigma_t^2}\sum_{n=2}^N\left(\sin\theta_n-\sin\theta_1\right)^2,
\end{align}
and
\begin{align}
\label{J_xy_2}&J\left({\phi_n}\right)_{12} = J\left({\phi_n,\varphi_n}\right)_{21}\notag\\
= &\dfrac{1}{{2c^2\sigma_t^2}}\sum_{n=2}^N\left(\sin\theta_n-\sin\theta_1\right)\left(\cos\theta_n-\cos\theta_1\right).
\end{align}
Based on \eqref{fisher_2}, we express the covariance matrix of the true Eve's location as $\mathbf{V} = \mathbf{J}^{-1}$. We further define $\mathbf{V}$ as
\begin{align}
\label{v_pos_2} \mathbf{V} = \left[
\begin{array}{*{20}{c}}
  \sigma_{x}^2&\sigma_{xy}\\
  \sigma_{yx}&\sigma_{y}^2
\end{array}\right],
\end{align}
where $\sigma_{xy}=\sigma_{yx}$. We denote the estimated Eve's location by ${\xi}_E = \left[{x}_E,{y}_E\right]$, and the correlation coefficient by
$
\rho = \sigma_{xy}/\left({\sigma_x\sigma_y}\right).
$
As such, the distribution of the estimated Eve's location can be expressed as
\begin{align}
\label{p_xy} P({\xi}_E)&= \dfrac{1}{2\pi\sqrt{1-\rho^2}\sigma_x\sigma_y}\exp\left\{-\dfrac{1}{2\left(1-\rho^2\right)}\left(\dfrac{\left({x}_E-{x}_0\right)^2}{\sigma_x^2}\right.\right.\notag\\
&\left.\left.+\dfrac{\left({y}_E-{y}_0\right)^2}{\sigma_y^2}
-\dfrac{2\rho\left({x}_E-{x}_0\right)\left({y}_E-{y}_0\right)}{\sigma_x\sigma_y}\right)\right\}.
\end{align}

In order to examine the impact of the uncertainty in Eve's location, we adopt an ``average'' measure of $P_{\text{out}}\left(R_S\right)$, which is  given by
\begin{align}
\label{average_outage} \overline{P}_{\text{out}}\left(R_s\right) = \int_{-\infty}^{\infty}\int_{-\infty}^{\infty}{P}_{\text{out}}\left(R_s\right)P\left(\xi_E\right)dx_Edy_E.
\end{align}
We note that a closed-form expression for $\overline{P}_{\text{out}}\left(R_s\right)$ in \eqref{average_outage} is not attainable. As such, we will numerically evaluate the impact of Eve's location uncertainty on the secrecy outage probability in Section V. Specifically, such an evaluation will be performed through the following steps: (1) Obtain  Eve's estimated location by randomly selecting a position $\xi_E = [x_E, y_E]$ from the distribution given in \eqref{p_xy}. (2) Based on this estimated  location, obtain the distance between Alice and  Eve's location $\hat{d}_E$, and the angle from Alice to the estimated Eve's location $\hat{\theta}_E$ (see  Fig.~\ref{system_model}). (3) Substitute $\hat{d}_E$ and $\hat{\theta}_E$ into \eqref{t1_result}, and obtain $P_{\text{out}}\left(R_S\right)$. (4) Repeat (1)-(3) and utilize all derived $P_{\text{out}}\left(R_S\right)$ in \eqref{average_outage},
thereby obtaining $\overline{P}_{\text{out}}\left(R_s\right)$.
\section{Numerical Results}
In this section we present numerical results to validate our analysis. Specifically, we first demonstrate the effectiveness of the proposed location-based beamforming scheme. We then examine in detail the impact
of the uncertainty in Eve's location on the secrecy performance of our proposed  scheme.
\begin{figure}[t!]
\begin{center}{\includegraphics[height=2.2in,width = 2.8in]{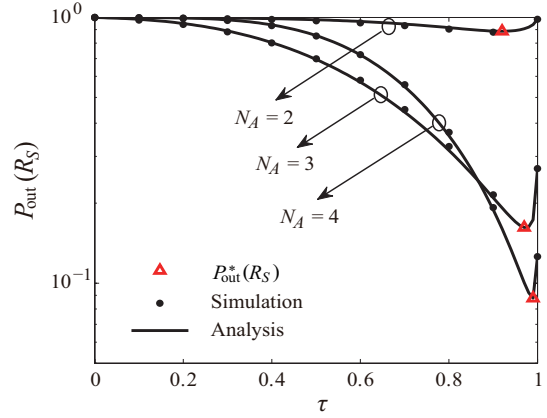}}
\caption{$P_{\text{out}}\left(R_S\right)$ versus $\tau$ for different values of $N_A$ with $N_E=2$, $K_B=10$ dB, $K_E=5$ dB, $\overline{\gamma}_B=\overline{\gamma}_E= 10$ dB, $\theta_B=\pi/3$, $\theta_E = \pi/4$, and $R_S=1$ bits/s/Hz.}\label{fig_side_a}
\end{center}
\end{figure}
\begin{figure}[t!]
\begin{center}{\includegraphics[height=2.2in,width = 2.8in]{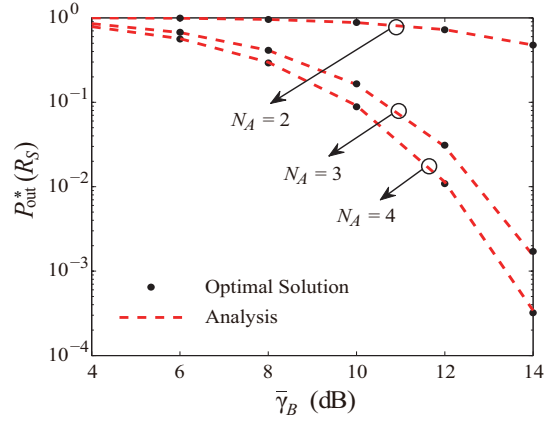}}
\caption{$P_{\text{out}}^{\ast}\left(R_S\right)$ versus $\overline{\gamma}_B$ for different values of $N_A$ with $N_E=2$, $K_B=10$ dB, $K_E=5$ dB, $\overline{\gamma}_E= 10$ dB, $\theta_B=\pi/3$, $\theta_E = \pi/4$, and $R_S=1$ bits/s/Hz.}\label{fig_side_b}
\end{center}
\end{figure}

In Fig. \ref{fig_side_a}, we plot $P_{\text{out}}\left(R_S\right)$ versus $\tau$ for different values of $N_A$ with $N_e=2$, $K_B=10$~dB, $K_E=5$~dB, $\overline{\gamma}_B=\overline{\gamma}_E= 10$~dB, $\theta_B=\pi/3$, $\theta_E = \pi/4$, and $R_s=1$ bits/s/Hz. We first observe that the analytical curves, generated from Proposition \ref{p1} and Theorem \ref{t1}, precisely match the simulation points marked by black dots, thereby demonstrating the correctness of our analysis for $P_{\text{out}}\left(R_S\right)$ in Theorem \ref{t1}. Second, we see that there exists a unique $\tau^{\ast}$ that minimizes $P_{\text{out}}\left(R_S\right)$ for each $N_A$. Third, we see that the minimal $P_{\text{out}}\left(R_S\right)$, denoted by $P_{\text{out}}^{\ast}\left(R_S\right)$, decreases significantly as $N_A$ increases. Furthermore, we observe that the optimal $\tau^{\ast}$ that achieves $P_{\text{out}}^{\ast}\left(R_S\right)$ approaches $1$ as $N_A$ increases. This reveals that the optimal beamforming vector $\mathbf{w}^{\ast}$ that minimizes $P_{\text{out}}\left(R_S\right)$ approaches $\mathbf{w}_{\text{ZF}}$ as $N_A$ increases.

In Fig. \ref{fig_side_b}, we plot $P_{\text{out}}^{\ast}\left(R_S\right)$ versus $\overline{\gamma}_B$ for different values of $N_A$. In this figure, we have adopted the same system configurations as those in Fig. \ref{fig_side_a}. The analytical curves, represented by red dashed lines, are generated from Proposition \ref{p1} and Theorem \ref{t1} with the optimal $\tau^{\ast}$ which minimizes $P_{\text{out}}\left(R_S\right)$ being selected for different values of $N_A$. The optimal beamforming solutions, represented by `$\bullet$' symbols, are obtained from minimizing $P_{\text{out}}\left(R_S\right)$ via an exhaustive search (i.e., a full multi-dimensional search) for different values of $N_A$. We first see that the minimal secrecy outage probability $P_{\text{out}}^{\ast}\left(R_S\right)$ achieved by our proposed beamforming scheme is almost the same as the optimal beamforming solution found via exhaustive search. This shows the optimality of our proposed scheme. Second, we see that $P_{\text{out}}^{\ast}\left(R_S\right)$ decreases significantly as $N_A$ increase. This reveals that adding extra transmit antennas at Alice improves the secrecy of the adopted system. We further see that $P_{\text{out}}^{\ast}\left(R_S\right)$ monotonically decreases as $\overline{\gamma}_B$ increase. This reveals that the secrecy outage probability reduces when Alice uses a higher power to transmit.

\begin{figure}[t!]
\begin{center}{\includegraphics[height=2.2in,width = 2.8in]{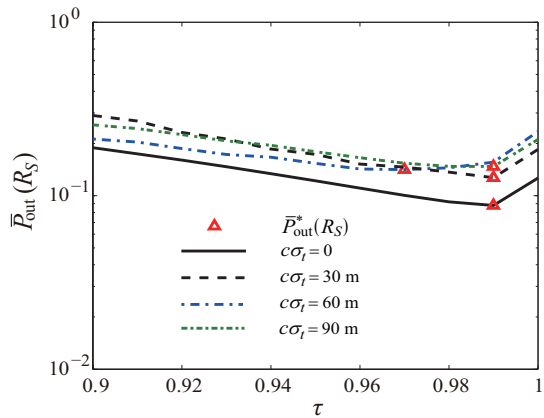}}
\caption{$\overline{P}_{\text{out}}\left(R_S\right)$ versus $\tau$ for different values of $c\sigma_t$ with $N_A=4$, $N_E=2$, $K_B=10$ dB, $K_E=5$ dB, $\overline{\gamma}_B=\overline{\gamma}_E= 10$ dB, $\theta_B=\pi/3$, $\theta_E = \pi/4$, and $R_S=1$ bits/s/Hz.}\label{fig_side_c}
\end{center}
\end{figure}

In Fig. \ref{fig_side_c}, we plot $\overline{P}_{\text{out}}\left(R_S\right)$ versus $\tau$ for different levels of Eve's location uncertainty using the procedures described in Section IV. The level of Eve's location uncertainty is represented by $c\sigma_t$. The larger $c\sigma_t$ is, the less accurate Eve's location is. In this figure, we consider that Alice and Bob are located in $[0~\text{m},~0~\text{m}]$ and $[1225~\text{m},~707~\text{m}]$, respectively. We also consider that the true location of Eve is $[1000~\text{m},~ -1000~\text{m}]$. For illustration purposes, we adopt $\eta = 4$. We see that there exists a unique $\tau^{\ast}$ that minimizes $\overline{P}_{\text{out}}\left(R_S\right)$ for each $c\sigma_t$. We also see that the minimal $\overline{P}_{\text{out}}\left(R_S\right)$ increases as $c\sigma_t$ increases, which demonstrates that the secrecy performance of our proposed beamforming scheme decreases, as the level of uncertainty in Eve's location increases. Although not completely shown here, we close by noting that our  results approach the appropriate solutions as the location uncertainty approaches both zero and infinity (i.e., location unknown), and show the expected trends between these two extremes.

\section{Conclusion}
In this work we have proposed a new location-based beamforming solution for Rician wiretap channels, in which a multi-antenna source communicates with a single-antenna receiver in the presence of a multi-antenna eavesdropper. In our scheme, we assumed that the CSI from the legitimate receiver is  known at the source, while the only available information on the eavesdropper is her location. We  showed how the  beamforming vector that minimizes the secrecy outage probability of our scheme can be obtained via our simplified analytical expression for the secrecy outage probability.  We also examined the impact of the eavesdropper's location uncertainty on the secrecy performance, showing that secrecy can still exist over a wide range of (anticipated) location inaccuracies. The results presented here are of importance to a range of realistic wiretap channels in which the only information known on an eavesdropper is a noisy estimate of her location.

\end{document}